\documentclass[%
 aip,a4paper,onecolumn,
 amsmath,amssymb,
reprint
]{revtex4-1}
\usepackage{graphicx}
\usepackage{dcolumn}
\usepackage{bm}

\usepackage[T1]{fontenc}
\usepackage{mathptmx,xcolor}

\renewcommand{\[}{\begin{equation}}
\renewcommand{\]}{\end{equation}}
\def\bea{\begin{eqnarray}}
\def\eea{\end{eqnarray}}
\def\nn{\nonumber\\}

\newcommand{\B}{{\bf B}}

\newcommand{\PP}{{\hat \rho}}

\newcommand{\A}{{\bf A}}

\newcommand{\p}{{\bf p}}

\renewcommand{\v}{{\bf v}}
\renewcommand{\r}{{\bf r}}
\newcommand{\R}{{\bf R}}

\newcommand{\dda}{\partial_{\lambda_1}}
\newcommand{\ddb}{\partial_{\lambda_2}}

\newcommand{\equ}[1]{Eq.~(\ref{#1})}
\newcommand{\eqs}[2]{Eqs.~(\ref{#1}) and (\ref{#2})}

\def\bra#1{\langle#1\vert}
\def\ket#1{\vert#1\rangle}
\def\ev#1{\langle#1\rangle}
\def\me#1#2#3{\langle#1| \, #2 \, |#3\rangle}
\def\runtime{(\the\time)\qquad\the\month/\the\day/\the\year}
\def\today
 {\count10=\year\advance\count10 by -2000 \number\day--\ifcase
  \month \or Jan\or Feb\or Mar\or Apr\or May\or Jun\or
             Jul\or Aug\or Sep\or Oct\or Nov\or Dec\fi--\number\count10}

\def\hour{\count10=\time\count11=\count10
\divide\count10 by 60 \count12=\count10
\multiply\count12 by 60 \advance\count11 by -\count12\count12=0
\number\count10 :\ifnum\count11 < 10 \number\count12\fi\number\count11}
\begin{document}

\title{{\bf Molecular Berry curvatures and the adiabatic response tensors}}

\date{\sf DRAFT: run through \LaTeX\ on \today\ at \hour}
\author{Raffaele Resta}
 \email{resta@iom.cnr.it}
\affiliation{ 
Istituto Officina dei Materiali IOM-CNR, Strada Costiera 11, 34151 Trieste, Italy
}%

 \homepage{http://www-dft.ts.infn.it/\~{}resta/.}
\affiliation{Donostia International Physics Center, 20018 San Sebasti{\'a}n, Spain
}%

\begin{abstract} 
Adiabatic transport in a many-electron system is expressed in terms of the appropriate Berry curvature, owing to the Niu-Thouless theory [J. Phys A {\bf 17}, 2453 (1984)]; the main equation is very compact and very general. I address here three paradigmatic adiabatic response tensors----the atomic polar tensor, the atomic axial tensor, and the rotational $g$ factor---and I show that, for all of them, the known formulas do not need an independent proof. They are just case studies of the general expression, for different choices the curvature's two arguments.
\end{abstract}

\maketitle


\section{Introduction}

The quantity which today is commonly addressed as a Berry curvature was introduced in a milestone 1984 paper by Michael Berry;\cite{Berry84} at about the same time, Niu and Thouless (NT) established the theory of adiabatic charge transport in a many-electron system in terms of the same quantity.\cite{Niu84} 
The curvature is a quasi-static quantity, in the sense that its definition requires solely the instantaneous ground eigenstate; yet it encodes the lowest-order effect of the excited states on the adiabatic evolution. 

While the notion of Berry curvature has permeated several areas of condensed matter physics,\cite{rap_a20,Xiao10,Vanderbilt} its role in molecular physics is not generally appreciated.  I am going to show here that the known expressions for some molecular adiabatic response tensors are indeed paradigmatic manifestations of the Berry curvature. When addressed by means of the NT theory of adiabatic evolution, such tensors become simple case studies. Actual derivation of the known formulas requires no independent proof: it is enough to plug the definition of each tensor into the general NT expression for the adiabatic evolution, based on the appropriate Berry curvature. I am going to show how this works in three cases: the atomic polar tensor, the atomic axial tensor, and the gyromagnetic factor. All three of these tensors  are a measure of how much electronic charge is ``dragged'' by the nuclear adiabatic motion, and this is precisely what the Berry curvature allows casting in a very compact formalism. It is also worth remarking that, when the nuclear motion is harmonic at frequency $\omega$, the NT time-evolution of the many-electron system is exact to linear order in the adiabaticity parameter $\omega$.

In Sec. \ref{sec:funda} I give the general definition of a Berry curvature and I detail its most basic features. In Sec. \ref{sec:adia} I consider a generic time-independent operator and the evolution of its expectation value over a given state when such state is adiabatically varied in time: the NT theory---rooted in a Berry curvature---provides a universal formula for such evolution. In Sec. \ref{sec:tensors} I work out in detail the three case studies mentioned above. A Conclusions Section is not contemplated; in the Appendix I deal with the case where the ground eigenstate is a Slater determinant of doubly occupied one-electron orbitals.

\section{The Berry curvature: fundamentals} \label{sec:funda}

Let $\hat{H}$ be a time-independent Hamiltonian and let $\ket{\Psi_n}$ its eigenstates with eigenvalues $E_n$; $\hat{H}$ depends on a couple of generic real parameters $(\lambda_1,\lambda_2)$: the eigenstates and eigenvalues are parameter-dependent as well. We assume the ground state to be nondegenerate for all $(\lambda_1,\lambda_2)$. 
The Berry curvature is by definition\cite{rap_a20,Xiao10,Vanderbilt} \bea \Omega(\lambda_1,\lambda_2) &=& i ( \ev{\dda \Psi_0 | \ddb \Psi_0} - \ev{\ddb \Psi_0 | \dda \Psi_0} ) \nn  &=& - 2 \,\mbox{Im } \ev{\dda \Psi_0 | \ddb \Psi_0} . \label{curva}\eea The parameters may have various physical interpretations and different dimensions; the curvature has the inverse dimensions of the product $\lambda_1 \lambda_2$. For a macroscopic homogeneous system $\Omega(\lambda_1,\lambda_2)$ is an extensive quantity.

The curvature admits a sum-over-states formula, first displayed in the original Berry's paper;\cite{Berry84}
\[ \Omega(\lambda_1,\lambda_2) = -2 \mbox{Im}  \sum_{n \neq 0} \frac{\ev{\Psi_0 | \dda\hat{H} | \Psi_n }\ev{\Psi_n | \ddb\hat{H} | \Psi_0 }}{(E_0 - E_n)^2} . \label{sum} \]  The formula perspicuously shows that the Berry curvature encodes---to lowest order---the effect of the excited states on the ground state when the quantum system is transported in the parameter space. \equ{sum}
has also the virtue of showing that the curvature becomes ill defined whenever the ground state becomes degenerate with the first excited state. 

In the special case where $\ket{\Psi_0}$ is the Slater determinant of doubly occupied single-particle orbitals---either Hartree-Fock or Kohn-Sham---$\Omega(\lambda_1,\lambda_2)$ becomes a sum of single-particle curvatures: the expression is presented in Appendix \ref{sec:single}.

The curvature is a geometrical quantity defined by the evolution of the ground-state projector \[ \PP = \ket{\Psi_0}\bra{\Psi_0} \label{proj}\] in the parameter space, and can equivalently be expressed directly in terms of $\PP$:  \[ \Omega(\lambda_1,\lambda_2) = - i \, \mbox{Tr } \{ \PP \,[ \,\dda \PP, \ddb \PP\,] \} , \label{prcurv} \] where ``Tr'' indicates the trace over the Hilbert space. The form of \equ{prcurv} explicitly displays the gauge-invariance of $\Omega$.
A gauge-transformation modifies $\ket{\Psi_0}$ by a phase factor, ergo leaves $\PP$ and the curvature unchanged. 

An important caveat is in order: the curvature is {\it not} gauge-invariant when the parameter $\lambda_1$ is identified with a macroscopic magnetic field $\B$, which is not a ``normal'' parameter on which the Hamiltonian explicitly depends. When in \equ{prcurv} $\lambda_1$ is identified with $\B$, the relevant entry is $\partial_{\B}\PP$, which actually is the functional derivative of $\PP$  with respect to the vector potential $\A(\r)$, times the derivative of $\A(\r)$ with respect to $\B$: the latter factor is manifestly gauge-dependent. 

Practical calculations are performed by projecting the Hamiltonian on a finite basis set, and are notoriously plagued by gauge-dependence problems. I am not discussing the issue in this work, ideally assuming a complete Hilbert space.

\section{Adiabatic transport} \label{sec:adia}

Let us suppose here that $\hat{H_t}$ is instead a time-dependent Hamiltonian, and $\ket{\Psi}$ one of its solutions: \[ \hat{H}_t \ket{\Psi} = i \hbar \ket{\dot\Psi} ; \] the time-dependent energy is \[ E(t) = \me{\Psi}{\hat{H}_t}{\Psi} . \label{energy} \]Let us also assume that a generic observable $\hat{O}$ can be written as the derivative of the Hamiltonian with respect to some parameter $\lambda_1$: \[ \hat{O} = \partial_{\lambda_1}  \hat{H}_t, \qquad \mbox{time independent} . \] The $\lambda_1$-derivative of \equ{energy} yields \[ \partial_{\lambda_1} E(t) =  \me{\Psi}{\hat{O}}{\Psi} + i\hbar \, (\, \ev{\partial_{\lambda_1}\Psi|\dot\Psi} - \ev{\dot\Psi|\partial_{\lambda_1}\Psi} \,) . \label{exact} \] 

Next suppose that the time-dependence of $\hat{H_t}$ owes to the time dependence of  a parameter $\lambda_2$ entering it: a nuclear coordinate in the cases of interest here. Upon replacing the exact time-dependent energy and state vector in \equ{exact} with the instantaneous ground-state eigenvalue and eigenvector one gets the adiabatic evolution of the ground-state observable in terms of the Berry curvature as \[ \ev{O(t)} = \me{\Psi_0}{\hat{O}}{\Psi_0} + \hbar \, \Omega(\lambda_1,\lambda_2)\, \dot\lambda_2(t) , \label{NT} \] where $\ket{\Psi_0}$ and the curvature depend implicitly on time. 
To the best of the author's knowledge, a similar formulation first appeared in the milestone NT paper.\cite{Niu84,Xiao10} In the following, for the sake of simplicity, I am going to call \equ{NT} the ``NT formula''.

When the Hamiltonian is time-reversal (T) invariant the ground state can be taken as real in the Schr\"odinger representation; its derivative with respect to a nuclear coordinate $\lambda_2$ is real as well. The observables of interest in the following are purely imaginary. Therefore the first term in \equ{NT} vanishes and the curvature, \equ{curva}, admits a simplified expression; the NT formula becomes in this special case \[ \ev{O(t)} = 2i\hbar \, \ev{\dda \Psi_0 | \ddb \Psi_0} \, \dot\lambda_2(t) , \qquad \mbox{T-invariant} \label{NT2}. \] In view of the rest of this paper, it is important to notice that \equ{NT2} becomes exact in the adiabatic limit, i.e.. when $\dot\lambda_2$ is infinitesimal. Ergo the expression \[ \frac{\partial \ev{O(t)}}{\partial \dot\lambda_2} = 2i\hbar \, \ev{\dda \Psi_0 | \ddb \Psi_0} = \hbar \, \Omega(\lambda_1,\lambda_2)\label{NT3} \] is also exact.

\section{The adiabatic response tensors} \label{sec:tensors}

\subsection{The electronic Hamiltonian}

Within the Born-Oppenheimer (also called Born-Huang) approximation one considers a bounded system of $N$ interacting electrons. The many-body Hamiltonian is \[ \hat{H}  = \frac{1}{2m} \sum_{i=1}^N \left({\bf p}_i + \frac{e}{c} \A(\r_i) \right)^2 + \hat{V} , \label{H}\]  where the potential $\hat{V}$ comprises one-body and two-body terms; it is a multiplicative operator in the Schr\"odinger representation, and is a function of the instantaneous nuclear positions $\{\R_s\}$.  We assume a singlet ground state and we neglect irrelevant spin variables. The Hamiltonian is T-invariant at $\A=0$.

It is also expedient to define the two operators: \[  \hat\v = \frac{1}{m}\sum_{i=1}^N \p_i , \qquad \hat{{\bf L}} =  \frac{1}{m}\sum_{i=1}^N \r_i\times \p_i : \label{pva} \]  velocity and angular momentum of the many-body system, respectively (at $\A=0$). In the following of this work Greek subscripts are Cartesian indices, and the sum over repeated indices is implicitly understood.

\subsection{Atomic polar tensor} \label{sec:aat}

When a given ion is transported with velocity $\v_s = \dot\R_s$, the induced electrical current is \[ {\bf J}_{s} = e Z_s {\bf v}_{s} - e \ev{{\bf v}(t)} ; \] the dimensionless atomic polar tensor (Born charge in solids and liquids) is then defined as \[ Z^*_{s,\alpha\beta} = \frac{1}{e} \partial_{v_{s\beta}} J_{s\alpha} =  Z_s \delta_{\alpha\beta} - \partial_{v_{s\beta}} \ev{{v}_\alpha(t)}. \label{j} \] 
Once this definition adopted, its microscopic expression immediately follows. In the NT formula it is enough to identify $\lambda_1$ with  $A_\alpha$ and $\lambda_2$ with $R_{s\beta}$;  the velocity is \[ \hat{{\bf v}} = \frac{c}{e}\partial_{{\bf A}} \hat{H} . \] The atomic polar tensor is then \[ Z^*_{s,\alpha\beta} = Z_s\delta_{\alpha\beta} + \frac{\hbar c}{e} \Omega(A_\alpha,R_{s\beta}) . \label{z2} .\] This equation is well known in molecular physics since the 1980s.\cite{Stephens85,Stephens87,Buckingham87,Barron} 

What I deem new here is the very concise derivation, as well as the realization that \equ{z2} is arguably the most straightforward case study of the very general NT formula, \equ{NT3}. \equ{z2} is gauge-invariant and invariant by translation of the coordinate origin.

\subsection{Atomic axial tensor} \label{sec:apt}

Let us suppose again that the nuclei are transported with velocities $\v_s$. The magnetic moment of the molecule is, in Gaussian units, \[  {\bf m} =\frac{e}{2c} \left(\sum_s Z_s  \R_s \times {\bf v}_s - \ev{{{\bf L}}(t)} \right)\label{m} .\] A common coordinate origin must be set for the two terms in parenthesis, and---for an arbitrary velocity distribution---the value of ${\bf m}$  depends on the coordinate origin.

The atomic axial tensor of nucleus $s$ is defined as \[ M_{s,\alpha\beta} = \frac{\partial m_\beta}{\partial v_{s\alpha}} = \frac{e}{2c} [\, \varepsilon_{\alpha\beta\gamma} Z_s  R_{s\gamma} - \partial_{v_{s\alpha}} \ev{{L}_\beta(t)} \,]. \label{aat} \]   Next, in order to exploit the NT formula, \equ{NT3}, all what is needed is recognizing the operator $\hat{L}_\beta$ as the Hamiltonian derivative with respect to some parameter. By adopting the central gauge centered on the momentum origin the parameter is clearly the $\B$ field: \[ \partial_{B_\beta} \hat{H} = \frac{e}{2c} \hat{L}_\beta . \label{Bder}\]  

Upon identifying $\lambda_1$ with $B_\beta$ and $\lambda_2$ with $R_{s\alpha}$, the NT formula of \equ{NT3} immediately yields the well known expression:\cite{Stephens85,Stephens87,Buckingham87,Barron} 
\bea  M_{s,\alpha\beta} &=& \frac{e}{2c} \varepsilon_{\alpha\beta\gamma} Z_s  R_{s\gamma} + \hbar \, \Omega(B_\beta,R_{s\alpha}) \nn &=& \frac{e}{2c} \varepsilon_{\alpha\beta\gamma}  Z_s  R_{s\gamma} + 2 i \hbar \ev{\partial_{B_\beta}\Psi_0|\partial_{R_{s\alpha}}\Psi_0} . \label{aat2} \eea   In the present straightforward derivation it appears as one of the numerous manifestations of the Berry curvature in adiabatic phenomena.

In order to arrive at the usual compact notation for the Berry curvature\cite{rap_a20,Xiao10,Vanderbilt}
one casts the antisymmetric tensor  $M_{s,\alpha\beta}$ into its equivalent vector form \[ \tilde{M}_{s\gamma} = \frac{1}{2}  \varepsilon_{\gamma\alpha\beta} M_{s,\alpha\beta} : \] \[ \tilde{{\bf M}}_s = \frac{e}{2c}  Z_s \R_s +i\hbar \, \ev{\partial_{\B} \Psi_0 | \times | \partial_{\R_s} \Psi_0}  . \label{aat3} \]  The magnetic moment linearly induced by the displacement of the $s$-th nucleus is then \[ {\bf m}_s = \tilde{{\bf M}}_s \times {\bf v}_s . \]

\eqs{aat2}{aat3} are not invariant by translation of the coordinate origin; it was early recognized that such ambiguity does not affect the measurable properties. They are also gauge-dependent; it is nonetheless known how to express the physical observables in gauge-invariant form.\cite{Stephens87} 

\subsection{Rotational $\bf g$ factor} \label{sec:g}

In order to simplify the algebra, we fix here the rotation axis, taken as the $z$-axis. Suppose a molecule is rigidly rotating around its center of mass; the rotational $g$ factor is defined as the ratio between the total magnetic moment of the molecule and its mechanical angular moment, expressed in dimensionless units.

If $\dot\vartheta$ is the angular velocity, the mechanical moment is \[ {\cal M}_z = \sum_s M_s R_s^2 \dot\vartheta ,\] where $R_s$ is the distance of the $s$-th nucleus from the axis and $M_s$ is its mass. The magnetic moment, analogously to \equ{m}, is \[ m_z = \frac{e}{2c}\left(\sum_s Z_s R^2_s \dot{\vartheta}  - \ev{L_z(t)} \right)  . \label{g} \] The $g$ factor can be cast as \[ g = \frac{1}{\mu_{\rm B}}\frac{m_z}{{\cal M}_z } = \frac{1+\sigma}{\mu_{\rm B}} \;\frac{\sum_s Z_s R^2_s}{\sum_s m_s R^2_s}  , \] where $\mu_{\rm B}$ is the Bohr magneton \[\mu_{\rm B}=  \frac{e\hbar}{2m c} . \] The magnetic screening $\sigma$ measures how much the electrons are ``left behind'' by the rotation of the nuclei. For $\sigma=0$ the electrons do not contribute and the nuclei rotate as naked, while for $\sigma=-1$ the electronic screening is complete and the molecule rotates as an uncharged rigid body. 

From \equ{g} the screening factor is \[ \sigma = - \frac{\partial_{\dot\theta} \ev{L_z(t)} }{\sum_s Z_s R^2_s} . \] By adopting a gauge centered on the rotation axis, and exploiting \equ{Bder}, the NT formula yields \[ \sigma = - \frac{2\hbar c}{e} \frac{\Omega(B_z,\vartheta)}{\sum_s Z_s R^2_s} , \label{sigma} \] where the Berry curvature $\Omega(B_z,\vartheta)$ is indeed a  constant (evaluated at $\B=0$, and $\vartheta$-independent). 

To the best of the author's knowledge, $\sigma$ was first recognized as a Berry curvature in Ref. \onlinecite{Ceresoli02}; the form of \equ{sigma} owes to Ref. \onlinecite{Stengel18}. \equ{sigma} can be cast into  a more familiar sum over states by means of \equ{sum}.

\appendix 
\section{Independent-electron Berry curvature} \label{sec:single}

In the mean-field case (either Hartree-Fock or Kohn-Sham) the singlet ground state is a Slater determinant of doubly occupied single-particle orbitals $\ket{\psi_j}$ with eigenvalues $\epsilon_j$. We assume that the parameters do not affect the two body terms in the Hamiltonian, ergo the $\lambda$-derivatives entering \equ{sum} are one-body operators: \[ \dda \hat{H} = \sum_{i=1}^N O_1(\r_i) , \] and analogously for $\ddb \hat{H}$. By the Slater-Condon rules the matrix elements in \equ{sum} are converted into one-body elements of $O_1$ and $O_2$ between occupied and unoccupied orbitals; the energy differences are easily expressed as well, and \equ{sum} becomes \[ \Omega(\lambda_1,\lambda_2) = 2\,i \hspace{-0.5cm}\sum_{\stackrel{j=\mbox{occupied}}{j'=\mbox{unoccupied}}} \hspace{-0.5cm}\frac{\ev{\psi_j | O_1| \psi_{j'}  }\ev{\psi_{j'} | O_2 | \psi_j}}{(\epsilon_j - \epsilon_{j'})^2} + \mbox{cc}  , \label{sum2} \] where the factor of two accounts for spin, and ``cc' stays for complex conjugate. Straightforward manipulations yield the independent-electron  version of \equ{curva}  as \[ \Omega(\lambda_1,\lambda_2) = 2 \, \tilde\Omega(\lambda_1,\lambda_2) , \]  \[ \tilde\Omega(\lambda_1,\lambda_2) = -2\,\mbox{Im } \sum_j \theta(\mu - \epsilon_j) (\, \ev{\dda \psi_j | \ddb \psi_j} \,) ,\] where $\mu$ is the Fermi level. 

$\tilde\Omega$ can be expressed in terms of the one-body ground-state projector \[ P = \sum_j \theta(\mu - \epsilon_j) \ket{\psi_j}\bra{\psi_j} : \] \[ \tilde\Omega(\lambda_1,\lambda_2) = - i \, \mbox{Tr } \{ P\,[ \,\dda P, \ddb P \,] \} .\] This shows that $\tilde\Omega$ enjoys a generalized form of gauge invariance: it is in fact invariant by any unitary mixing of the occupied orbitals between themselves.

\section*{Acknowledgments}
I have thoroughly discussed some key points of the present paper with Stefano Baroni, Ivo Souza, Massimiliano Stengel, and David Vanderbilt.
Work supported by the Office of Naval Research (USA) Grant No. N00014-20-1-2847.

\end{document}